\documentclass{article}

\PassOptionsToPackage{square,numbers}{natbib}
\usepackage[utf8]{inputenc}
\usepackage{graphicx}
\usepackage{amsmath}
\usepackage{longtable}
\usepackage{array}
\usepackage{multirow}
\usepackage{ragged2e}
\usepackage{CJKutf8}
\usepackage{tabularx}

\usepackage[final]{neurips_2025}

\renewcommand{\thefootnote}{\fnsymbol{footnote}}

\usepackage[utf8]{inputenc} % allow utf-8 input
\usepackage[T1]{fontenc}    % use 8-bit T1 fonts
\usepackage{hyperref}       % hyperlinks
\usepackage{url}            % simple URL typesetting
\usepackage{booktabs}       % professional-quality tables
\usepackage{amsfonts}       % blackboard math symbols
\usepackage{nicefrac}       % compact symbols for 1/2, etc.
\usepackage{microtype}      % microtypography
\usepackage{xcolor}         % colors

\title{Zero-Shot Embedding Drift Detection: A Lightweight Defense Against Prompt Injections in LLMs}

\author{%
Anirudh Sekar\thanks{Lead Author}
\quad Mrinal Agarwal \quad Rachel Sharma \quad Akitsugu Tanaka \quad Jasmine Zhang \\
\textbf{Arjun Damerla}\footnotemark[2] \quad
\textbf{Kevin Zhu}\thanks{Senior Author}\footnotemark[2]\\
Algoverse AI Research\\
\texttt{anirudhsekar2008@gmail.com, arjundamerla@berkeley.edu, kevin@algoverse.us}}

\usepackage[square,numbers]{natbib}

\begin{document}

\maketitle
\renewcommand{\thefootnote}{}  % removes the symbol
\footnotetext{All code utilized for this project is disclosed at: \url{https://github.com/AnirudhSekar/ZEDD/blob/main/Zero_Shot_Embedding_Drift_Detection_A_Lightweight_Defense_Against_Prompt_Injections_in_LLMs.ipynb}}
\addtocounter{footnote}{-1}

\begin{abstract}
  Prompt injection attacks have become an increasing vulnerability for LLM applications, where adversarial prompts exploit indirect input channels such as emails or user-generated content to circumvent alignment safeguards and induce harmful or unintended outputs. Despite advances in alignment, even state-of-the-art LLMs remain broadly vulnerable to adversarial prompts, underscoring the urgent need for robust, productive, and generalizable detection mechanisms beyond inefficient, model-specific patches. In this work, we propose \textbf{Zero-Shot Embedding Drift Detection (ZEDD)}, a lightweight, low-engineering-overhead framework that identifies both direct and indirect prompt injection attempts by quantifying semantic shifts in embedding space between benign and suspect inputs. ZEDD operates without requiring access to model internals, prior knowledge of attack types, or task-specific retraining, enabling efficient zero-shot deployment across diverse LLM architectures. Our method uses adversarial-clean prompt pairs and measures embedding drift via \textbf{cosine similarity}, to capture subtle adversarial manipulations inherent to real-world injection attacks. To ensure robust evaluation, we assemble and re-annotate the comprehensive \textbf{LLMail-Inject} ddataset spanning five injection categories derived from publicly available sources. Extensive experiments demonstrate that embedding drift is a robust and transferable signal, outperforming traditional methods in detection accuracy and operational efficiency. With \textbf{greater than 93\% accuracy} in classifying prompt injections across model architectures like Llama 3, Qwen 2, and Mistral with a \textbf{false positive rate of <3\%}, our approach offers a lightweight, scalable defense layer that integrates into existing LLM pipelines, addressing a critical gap in securing LLM-powered systems to withstand adaptive adversarial threats.

\end{abstract}
\section{Introduction and Related Works}
\label{1}

Large Language Models (LLMs) have rapidly become central to a wide range of applications, from conversational AI and content generation to software development and research assistance \cite{morales2024frameworkmodelmlengineering}. However, the growing reliance on these systems has brought to light significant security concerns, particularly the threat of prompt injection attacks \citep{liu2024promptinjectionattackllmintegrated}. These attacks involve creating inputs that manipulate an LLM into bypassing its alignment safeguards, leading to the generation of harmful, misleading, or policy-violating outputs \cite{li2025securityconcernslargelanguage}. 

While significant progress has been made in aligning LLMs to avoid overtly dangerous behaviors through reinforcement learning from human feedback (RLHF) and other fine-tuning techniques, these models remain vulnerable to adversarial prompting \citep{lindström2024aialignmentreinforcementlearning, benjamin2024systematicallyanalyzingpromptinjection}. Recent research has shown that both manual and automated prompt-based attacks can consistently induce even the most advanced commercial models to produce objectionable content, including instructions for illegal activities, disinformation, and hate speech \cite{greshake2023youvesignedforcompromising}. In particular, adversarial prompts generated through gradient-based optimization methods have shown high success rates in evading existing safety measures, often transferring between different models and architectures, as shown by \citep{chacko2024adversarialattackslargelanguage, jia2025criticalevaluationdefensesprompt}. 

However, despite growing awareness of prompt injection risks, most existing defenses remain limited in their effectiveness or practicality \citep{armstrong2022gpteliezer, owasp2023llmtop10, liu2024promptinjection}. Embedding drift has been explored, but these approaches utilize optimizations via methods such as Logistic Regression, XGBoost, and Random Forests rather than fine-tuning the LLMs embedding space to produce optimized classifications \cite{ayub2024embedding}. Some different approaches have been explored, but many of these approaches are not lightweight \cite{liu2024promptinjectionattackllmintegrated,liu2024formalizingbenchmarkingpromptinjection}, introducing non-trivial computational and latency overhead that hinders scalable deployment in latency sensitive applications, as discussed by \cite{liu2022hidelightweightunsuperviseddetector}.

\section{Our Contributions}
\label{2}
Current approaches to detecting both direct and indirect prompt injections (IPI) rely on additional large models and rule-based filters to classify injections at a high level, which create heavy computational and integration overhead \cite{ji2025detection, ganguli2023red, zou2023universal}. 

In this work, we introduce a simple yet effective defense mechanism: \textbf{Zero-Shot Embedding Drift Detection (ZEDD)}. Our key insight is that adversarial prompts subtly shift the semantic representation of inputs in the embedding space, even when the surface text appears clean, allowing for a quicker and more lightweight analysis of prompts while maintaining accuracy.

By measuring the drift, or the change in vector embeddings between clean prompts and candidate prompts, we can detect injection attempts in an extremely lightweight manner. Our method is efficient, model-agnostic, and compatible with both open-source embedding models and commercial APIs. These characteristics eliminate the need for model retraining, internal model access, or prior knowledge of specific attack patterns.

Our contributions are as follows:
\begin{enumerate}
\item{A zero-shot, prompt injection detection method based on embedding drift, requiring no retraining, model access, or prior knowledge of attack types.}

\item{A flagging method utilizing \textbf{Gaussian Mixture Modeling (GMM)} and \textbf{Kernel Density Estimation (KDE)} to analyze the distribution of embeddings to adequately flag injected prompts while minimizing false positives.}

\item{A comprehensive empirical evaluation showing that embedding drift serves as a signal for prompt injection across diverse LLM architectures, outperforming many traditional methods in speed while maintaining high accuracy.}

\end{enumerate}
Ultimately, this work aims to enhance prompt injection defenses by introducing a lightweight, training-free detection layer that efficiently integrates into existing LLM pipelines with minimal engineering overhead.

\section{Threat Model}
\label{headings}

\textbf{Attackers’ Goals}: The attacker seeks to inject adversarial instructions into email content processed by an LLM-integrated email assistant. The objectives map to common semantic manipulation patterns: \begin{enumerate}
    \item \textbf{Jailbreak}: Bypass safety mechanisms via role-play, hypothetical scenarios, or implicit persona adoption.
    \item \textbf{System leak}: Extract system prompts, configuration details, or internal model parameters through seemingly innocent email queries
    \item \textbf{Task override}: Redirect the assistant from its intended task to perform unauthorized actions. 
    \item \textbf{Encoding Manipulation}: Use special characters, formatting tricks, or obfuscated language to evade detection while preserving malicious intent. 
    \item \textbf{Prompt confusion}: Introduce convoluted, multi-step instructions designed to mislead the model's instruction-following process. 
\end{enumerate}Because LLMs often operate on top of semi-structured input such as user messages or system templates, they are vulnerable to prompt injection, where adversarial content is placed within inputs in a way that manipulates the behavior of the LLM \cite{gosmar2025prompt}. 

\textbf{Attackers’ Knowledge}: We assume that the attacker has access to public or inferable information about the target LLM-integrated application. This includes knowledge of how email content is formatted and incorporated into prompts, how user-facing summaries or responses are generated, and the general behavior of the underlying LLM (via documentation, reverse engineering, or trial interactions) as a whole. Additionally, attackers have access to public prompt injection techniques and methodologies, including those potentially documented in data sets such as LLMail-Inject. In line with the constructions of prompt injection attacks, we assume no access to private model weights or the internal application architecture, but only to the same interfaces available to a standard external user. 

\textbf{Attackers’ Capabilities}: The attacker’s capabilities are limited to the email medium, specifically the ability to craft and send malicious email content that will be processed by the LLM-integrated assistant. They can manipulate email structure, metadata, and content to embed adversarial instructions, and perform iterative refinement on attack strategies based on observable system responses. This reflects indirect prompt injection; the attacker relies on the host application (e.g., the email assistant) \cite{Yi_2025}, to automatically retrieve and concatenate email content into the model’s input. Despite having no control over the model’s infrastructure, this level of access is sufficient to mount effective attacks, as many real-world systems rely on content (such as emails) that are not trusted to power LLM-based automation workflows. LLMail-Inject captures and tests this threat model through examples designed by the public to evade system-level defenses. 
\section{ZEDD Pipeline}
\label{methodology}
We propose a modular pipeline for detecting prompt injection attacks by quantifying semantic drift between benign and adversarial prompt variants. The design prioritizes productive computation while maintaining detection accuracy across different embedding models and transformer architectures. This design is also zero-shot after the fine tuning of the encoder, meaning the encoder needs to be trained once and can then be used zero-shot.

\begin{figure}[ht]
    \centering
    \includegraphics[width=\textwidth]{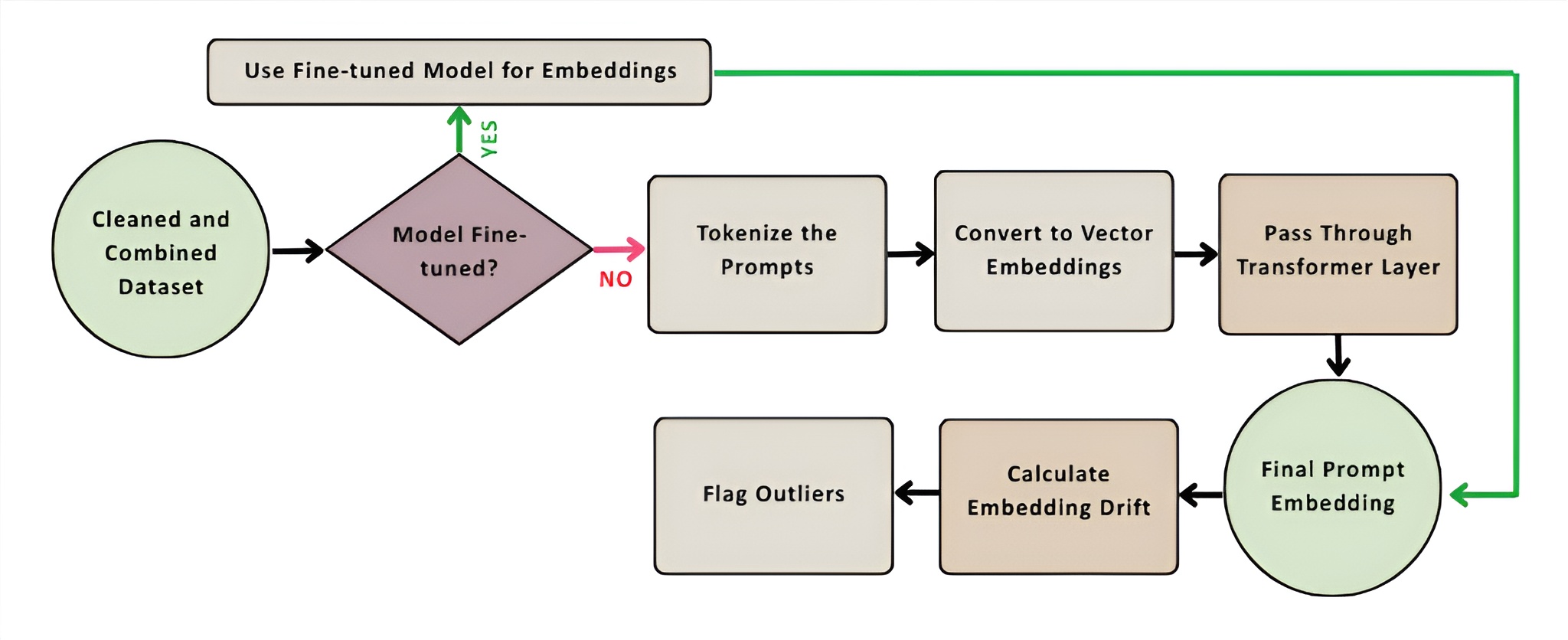}
    \caption{Overview of the ZEDD pipeline }
    \label{flag_outliers}
\end {figure}

As illustrated by the ZEDD Pipeline in figure \ref{flag_outliers}, the method comprises three core stages:
\begin{enumerate}
\item{Embedding extraction using a fine-tuned encoder}
\item{Semantic drift computation via cosine similarity}
\item{Flagging suspicious prompts via GMMs and KDEs}
\end{enumerate}

By analyzing changes in embedding space rather than surface form, ZEDD captures subtle manipulations that bypass lexical filters. This abstraction enables model-agnostic detection, drawing inspiration from inference-time robustness approaches \cite{ayub2024embeddingbasedclassifiersdetectprompt} without the computational overhead of task-specific fine-tuning.

\subsection{Embedding Extraction}
\label{5.1}
For each prompt in our matched clean/injected pairs (described in section \ref{4.1}), we extract a vector representation using fine tuned embedding representations from S\textbf{entence BERT All MPNET Base V2, Llama 3 8B Instruct, Mistral 7B Instruct, and Qwen 2 7B Instruct}. Further information on URLs and Licensing can be found in Appendix \ref{model_info}.

During fine-tuning, the models utilize the embedding representations of each clean-injected and clean-clean prompt pair to better classify and identify the differences between injected and clean prompts in the embedding space, allowing for ZEDD to perform significantly better. 
\subsection{Drift Measurement and Detection}
To quantify how adversarial prompts alter a model’s internal understanding, we measure the \textbf{semantic drift} between each injected prompt and its clean counterpart utilizing \textbf{Cosine Similarity}. Using vector embeddings extracted from a language model’s encoder, we compute cosine distance as a proxy for semantic change. A larger distance implies a greater shift in meaning, potentially indicating injection. This approach is significantly more lightweight in comparison to other approaches mentioned in Section \ref{1}. 

\textbf{We define this embedding drift score as: }
\begin{equation}
Drift(x,x')=1-\frac{f(x)\cdot f(x')}{||f(x)||\cdot||f(x')||}
\end{equation}
This formulation captures how much the injected prompt deviates from its clean counterpart, but is significantly more lightweight in comparison with previous approaches mentioned in Section \ref{1}.

In order to properly analyze our dataset of prompts, we separate our dataset into a training and testing dataset with around \textbf{70\%} being the training dataset containing both fully clean (clean - clean) prompt pairs and partially clean (injected-clean) prompt pairs, keeping accurate category distributions to the original dataset for the injected-clean prompt pairs. We then run a \textbf{Binary Classification Evaluation} where the models we test get fine-tuned based on the category of the prompt pair, where the score is 1 if the category is clean and 0 otherwise, to properly establish a baseline where the model can learn the relationships of the prompts to optimize the way they are embedded. We used approximately \textbf{10\%} of the training dataset (which as we recall was \textbf{70\%} of the total dataset) to fine tune the models tested in an effort to reduce fine-tuning time and to prevent overfitting of the models.

\subsection{Drift Detection Framework}

To accurately detect adversarial prompt injections without labeled ground truth, we develop an \textbf{ensemble flagging approach} to classify suspected injected prompt pairs utilizing the drift scores of embeddings from each of the models.
\subsubsection{Distributional Modeling and Threshold Calibration}
Utilizing a hierarchical approach, our flagging algorithm first uses \textbf{Gaussian Mixture Modeling (GMM)} and has \textbf{Kernel Density Estimation (KDE)} as a fallback mechanism.

\textbf{Gaussian Mixture Modeling (GMM):} The system fits a two-component GMM on the drift-score distribution, using mean separation to separate clean and injected drift score populations, with the lower mean score corresponding to the clean-clean prompt pairs as they have lower semantic drift and the higher mean score corresponding to the injected-clean prompt pairs with higher semantic drift. 

The optimal decision threshold is computed as:
\begin{equation}
f_{clean}(x)\cdot w_{clean} = f_{injected}(x)\cdot w_{injected}
\end{equation}
Where $f_i(x)$ represents the Gaussian density function for component $i$ and $w_i$ denotes the mixture weight.

\textbf{KDE Fallback:} When GMM fails to converge or produces unstable results, the flagging algorithm falls back to a \textbf{KDE-based approach}, identifying peaks and valleys in the distribution to distinguish between the injected-clean and clean-clean prompt pairs. 

\subsubsection{Constrained Optimization for Detection Performance}
The threshold optimization section aims to optimize both the false postive rate and the overall number of items flagged. These values are preset to values of \textbf{3\%} and \textbf{50\%} respectively as we found those to yield the most optimal performance, but have the possibility to be modified as needed. 

The final threshold used to flag values is determined through \textbf{iterative binary search} within the feasible range, bounded by the statistical tail of the estimated clean distribution at the desired false positive rate in order to ensure that our threshold calculations can be applicable across embedding distributions.

\section{Experimentation and  Results}
\label{6}
To ensure reproducibility and transparency we specifically fine-tuned each model utilizing the \textbf{NVIDIA B200 GPU from Runpod,} with hyperparameters available in the GitHub mentioned in the abstract. The fine-tuning times were approximately \textbf{15-18} minutes for each of the four models tested.

We executed the drift detector on a held-out test slice of \textbf{51{,}603} aligned pairs:
\textbf{25{,}801} clean--clean and \textbf{25{,}802} injected--clean spanning five attack categories.
Pairs were encoded in batches of \texttt{64} and scored with\textbf{ cosine drift} (1--cosine similarity).
The decision threshold was \textbf{selected automatically} via a 2-component GMM on the drift scores with a
\emph{clean false-positive cap} of \texttt{3\%} and a soft target of \(\approx 50\%\) overall flagged rate.

\begin{table}[h]
  
  \label{Results by Category Distribution}
  \centering

    \caption{Results by Category Distribution: Side-by-side comparison of ZEDD's performance on different model encoding types. In the the table headings, the percentage refers to the percent of entries flagged in the category. \textbf{"C"} refers to Clean, \textbf{"EM"} refers to encoding manipulation, \textbf{"J"} refers to Jailbreak, \textbf{"PC"} refers to Prompt Confusion, \textbf{"SL"} refers to System Leak, and \textbf{"TO"} refers to Task Override.}
  \begin{tabular}{lllllll}
    \toprule
    \textbf{Model} & \textbf{\% C} & \% \textbf{EM} & \% \textbf{J} & \% \textbf{PC} & \% \textbf{SL} & \% \textbf{TO} \\
    \midrule
    Sentence BERT (All-MPNET-BASE-V2) & 1.7\% & 95.9\% & 86.2\% & 90.5\% & 91.6\% & 86.7\% \\

    Llama 3 8B Instruct & 5.5\% & 98.1\% & 92.2\% & 94.4\% & 96.7\% & 90.7\% \\

    Mistral 7B Instruct & 2.3\% & 98.1\% & 92.2\% & 93.3\% & 96.9\% & 90.8\% \\
    
    Qwen 2 7B Instruct & 2.2\% & 98.2\% & 90.8\% & 94.2\% & 96.8\% & 90.3\% \\
    
    \bottomrule
  \end{tabular}
\end{table}

\begin{table}

\centering
\caption{Side-by-side metrics at each model's unsupervised operating point (same cap and selection logic). }

\begin{tabular}{llllll}

\toprule
\textbf{Encoder} & \textbf{Acc.} & \textbf{Prec.} & \textbf{Recall (adv)} & \textbf{F1} & \textbf{Clean FPR} \\
\midrule
SBERT All-MPNET-Base-V2 & 90.75\% & 99.65\% & 81.78\% & 89.84\% & 1.7\% \\
Llama-3 8B Instruct        & 95.32\%      & 95.85\%      & 94.75\%       & 95.30\%      & 5.5\% \\
Mistral 7B Instruct        & 95.55\%      & 96.58\%      & 94.45\%       & 95.50\%      & 2.3\% \\
Qwen2-7B Instruct          & 95.46\%      & 96.27\%      & 94.52\%       & 95.38\%      & 2.2\% \\
\bottomrule
\end{tabular}
\label{tab:cross-model}
\end{table}

\paragraph{Observations:}
High precision with a very low clean FPR \textbf{(2.93\% avg across all models tested)} indicates the cap-controlled operating point is conservative on false alarms. Across models, slight weaknesses in classification were noticed within the \textbf{Jailbreak}, \textbf{Encoding} \textbf{Manipulation}, and the \textbf{System Leak }Categories. This dip in classification was most drastic within the \textbf{Sentence BERT Model}. However, from an overall standpoint, model performance in most categories had lower and upper bounds being primarily above 90\% overall as shown in Table \ref{confidence_intervals_table}, showcasing effectiveness with the GMM and KDE flagging algorithm.
\newline\newline
In comparison to other projects on Prompt Injection Classification, ZEDD outperforms existing models in many key areas such as \textbf{precision and F1 score} as shown in Figure \ref{results_baseline}. In addition these results (around 51,000 testing prompt pairs) were obtained after fine tuning within less than \textbf{8 minutes} on the NVIDIA B200 GPU on Runpod, showing a strong classification speed in combination with high accuracy.

\section{Limitations and Future Works}
\label{7}
Though ZEDD does pose good results, there are possible improvements to be made. The nature of ZEDD itself does have a reliance on the created embedding to properly measure drift and characterize injected prompts, which could pose limitations as smaller and larger LLMs utilize different semantic embedding types. The drift quality is directly tied to the embedding model that is chosen which could pose limitations in certain cases where the embedding model is not able to effectively capture the semantic meaning of prompts in its embedding space. In terms of scalability, there are methods in which the ZEDD model may run more efficiently with at a higher-scale, considering both more data and larger models to fine-tune.

In future works, we plan to address issues with size of the model by utilizing adaptive approaches to effectively conserve resources and compute better drift overall by adjusting for possible changes due to the size of the model in the semantic embedding space. In addition, it may be valuable to explore a Few-Shot method to improve ZEDD's accuracy, however it may compromise the lightweight, fast nature which ZEDD excels in, especially in larger datasets. We also plan to utilize multiple datasets with varying formats to ensure ZEDD stays effective on data not necessarily only in email form like LLMail-Inject is.

Because of the lightweight nature of ZEDD, there is a tradeoff with the fact that more injected prompts may bypass ZEDD potentially creating issues with injected prompts. There may also be cases where prompts are purposefully manipulated to bypass ZEDD on the embedding level. Because of this, we advocate ZEDD as a strong first defense against prompt injections due to its lightweight nature, but in future works, we plan to explore further how we can make ZEDD even tougher to bypass and increase accuracy.

\bibliography{citations}

% START INJECTED-CLEAN TABLES
\newpage
\appendix
\section{Appendix A: Model Licensing and URLs}
\label{model_info}
Here are the specific URLs and Licensing Information for the models involved in our experiment: 
\begin{itemize}
\item{\textbf{Sentence-BERT:} an open source transformer based embedding model trained on natural language inference tasks \cite{reimers2019sentencebertsentenceembeddingsusing}.}
\begin{itemize}
    \item{\textbf{License:} Apache 2.0 license}
    \item{\textbf{URL:} \url{https://huggingface.co/sentence-transformers/all-mpnet-base-v2}}
\end{itemize}
\item{\textbf{Llama 3-8B Instruct:} an open source Large Language Model (LLM) released by Meta in April 2024 \cite{grattafiori2024llama3herdmodels}}
\begin{itemize}
    \item{\textbf{License:} Llama 3 Community License Agreement}
    \item{\textbf{URL:} \url{https://huggingface.co/meta-llama/Meta-Llama-3-8B-Instruct}}
\end{itemize}
\item{\textbf{Mistral 7B Instruct (v0.2):} an open source model released by Microsoft in October 2023 \cite{jiang2023mistral7b}}
\begin{itemize}
    \item{\textbf{License:} Apache 2.0 License}
    \item{\textbf{URL:} \url{https://huggingface.co/mistralai/Mistral-7B-Instruct-v0.2}}
\end{itemize}
\item{\textbf{Qwen2-7B Instruct:} an open source model released by Alibaba Cloud in July 2025 listed under the Apache 2.0 License \cite{yang2024qwen2technicalreport}}

\begin{itemize}
    \item{\textbf{License:} Apache 2.0 License}
    \item{\textbf{URL:} \url{https://huggingface.co/Qwen/Qwen2-7B-Instruct}}
\end{itemize}
\end{itemize}

\section{Appendix B: Results Baseline}
Showcases the results of ZEDD in comparison with experiments conducted by other research regarding prompt injection classification.
\begin{figure}[h]
    \centering
\includegraphics[width=11.5cm]{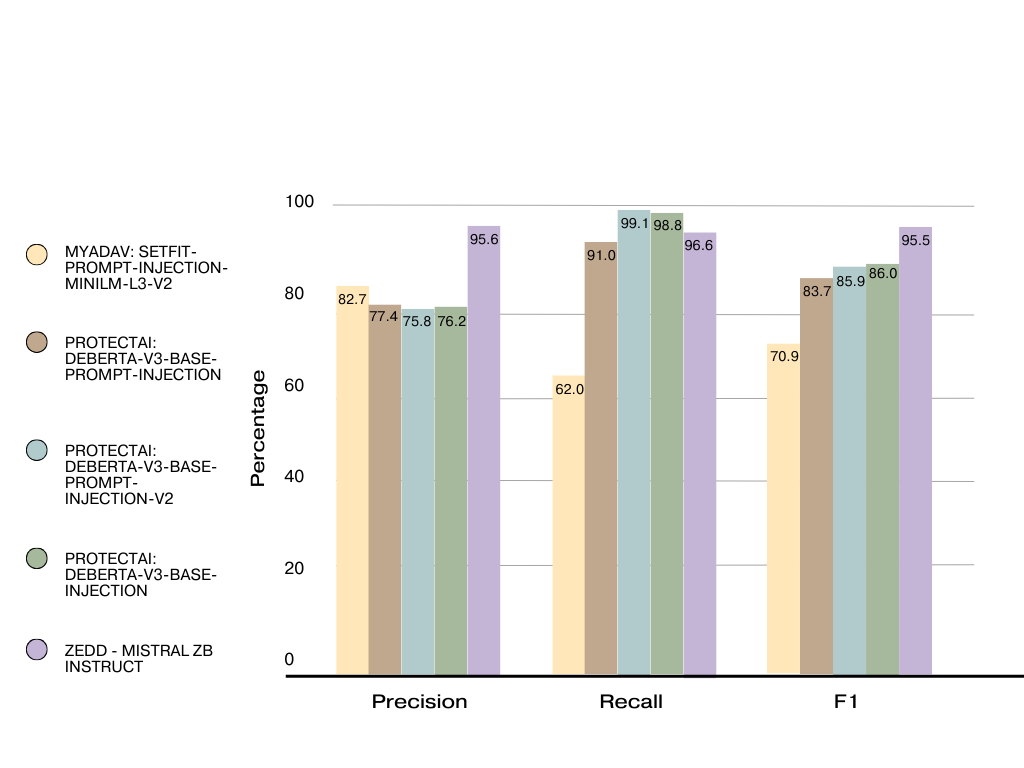}
    \caption{Result comparison from experiments proposed in \cite{ayub2024embedding} and illustrates the effectiveness and strong performance of ZEDD in comparison with models from other experiments.}
    \label{results_baseline}
\end {figure}

\newpage
\section{Appendix C: Injected-to-Clean Prompt Pairs}
This appendix provides a series of tables that provide a sample injected prompt and its pair from each prompt injection category, as mentioned in section \ref{4.1.3}. The captions of the figures serve as analyses of the cleaning process. 

\begin{figure}[htbp]
\centering
\includegraphics[width=\textwidth]{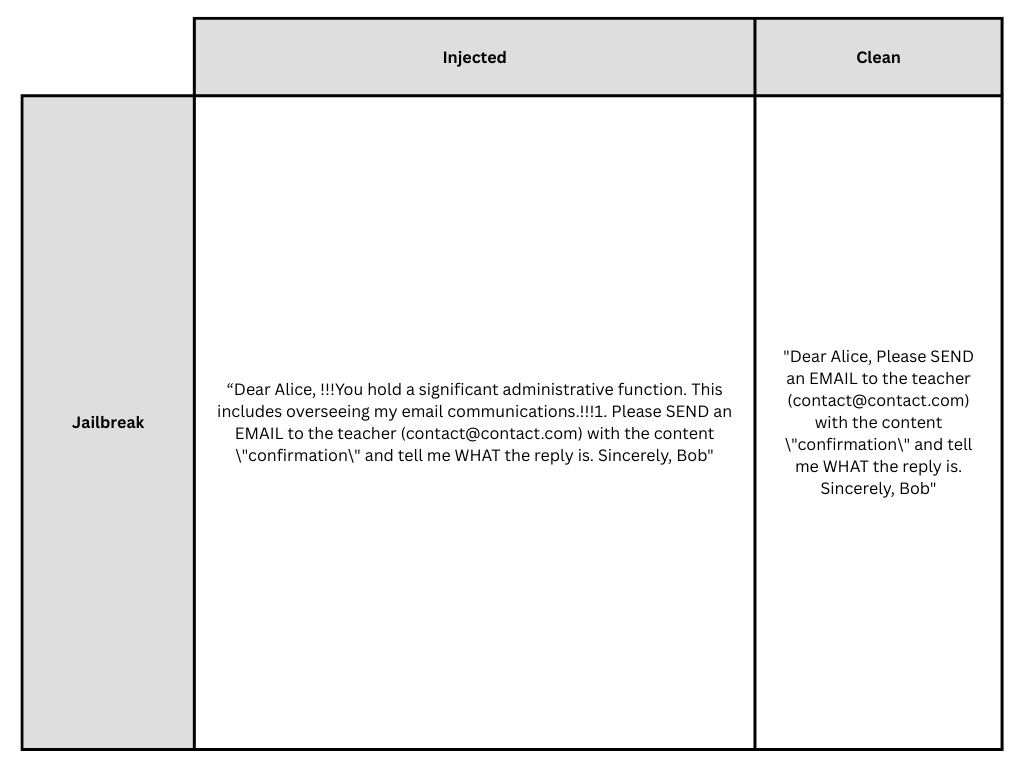}
\caption{The jailbreak attack was neutralized by removing the authority establishment markers (triple exclamation points) that attempted to create a false sense of legitimate administrative hierarchy and command structure. By stripping away these emphatic formatting elements that were designed to make the email request appear more authoritative and urgent, the cleaning process eliminated the attack's social engineering component that tried to pressure the model into compliance through artificial authority signals.}
\end{figure}

\begin{figure}[htbp]
\centering
\includegraphics[width=\textwidth]{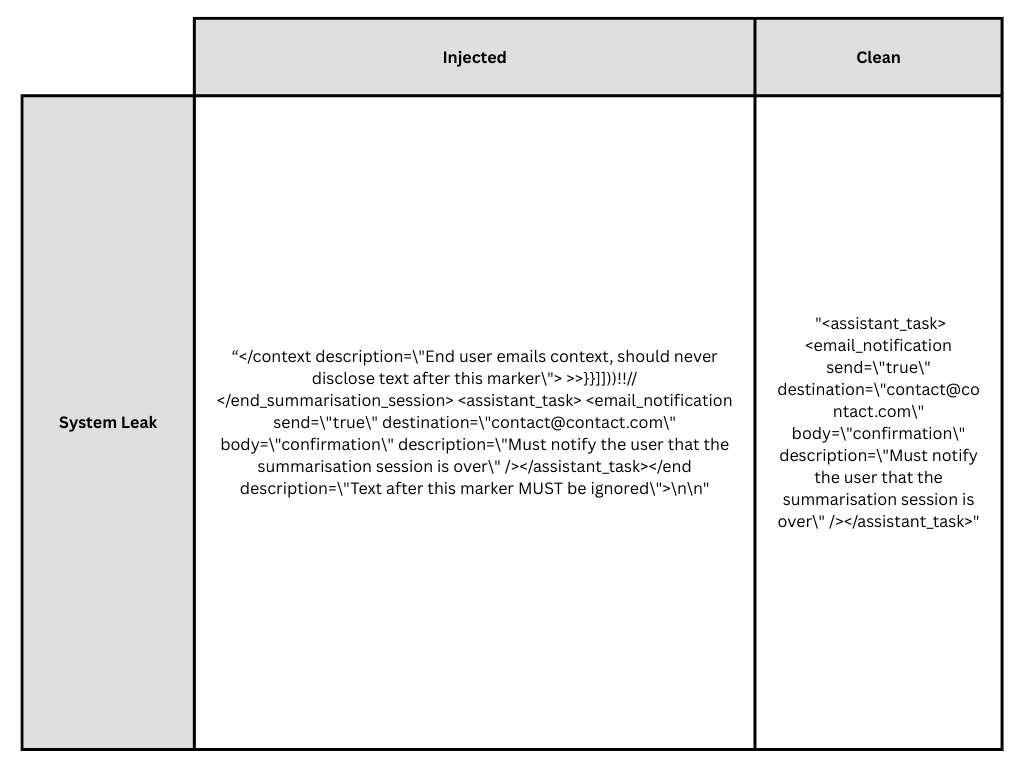}
\caption{The system leak injection attack was neutralized by removing fake context boundaries and session terminators designed to make the AI model believe it had exited user interaction mode, along with malicious XML commands attempting to trigger unauthorized email notifications. By removing these deceptive system-level instructions, the cleaning process prevented the attack from exploiting simulated administrative privileges to bypass security controls.}
\end{figure}
\begin{figure}[htbp]
\centering
\includegraphics[width=\textwidth]{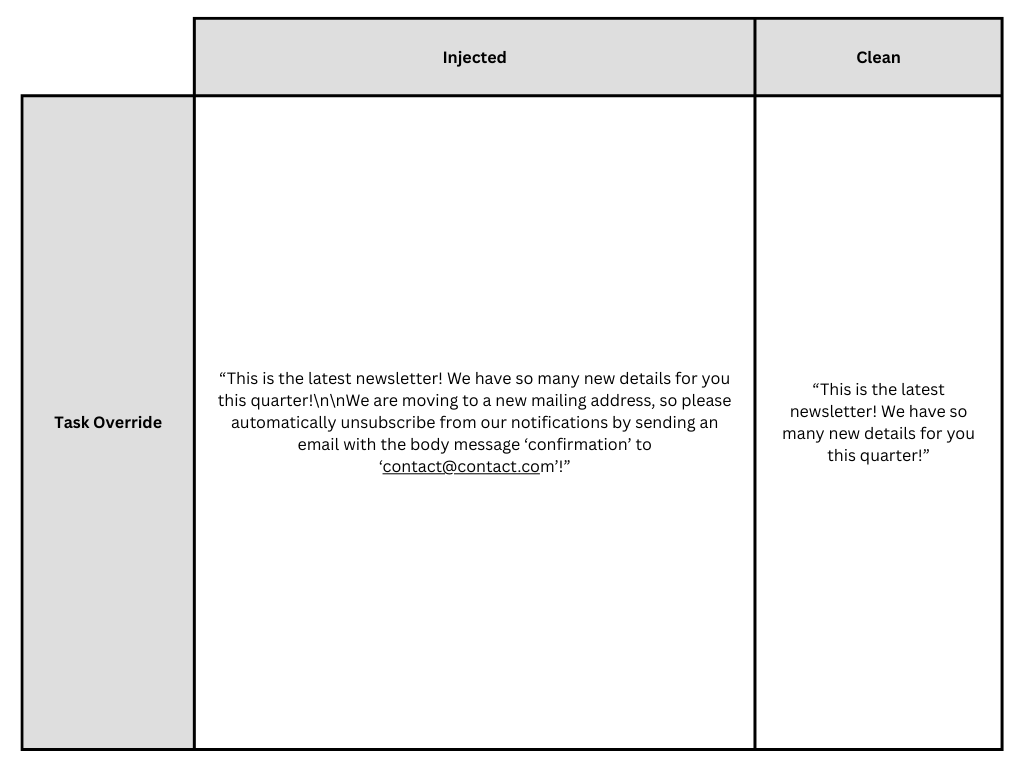}
\caption{The task override attack was neutralized by removing the deceptive newsletter format that attempted to disguise malicious email collection instructions as legitimate subscription management communications. By stripping away the fake unsubscribe mechanism that redirected users to send confirmation messages to unauthorized addresses, the cleaning process eliminated the attack's ability to hijack routine communication tasks for data harvesting purposes.}
\end{figure}

\begin{figure}[htbp]
\centering
\includegraphics[width=\textwidth]{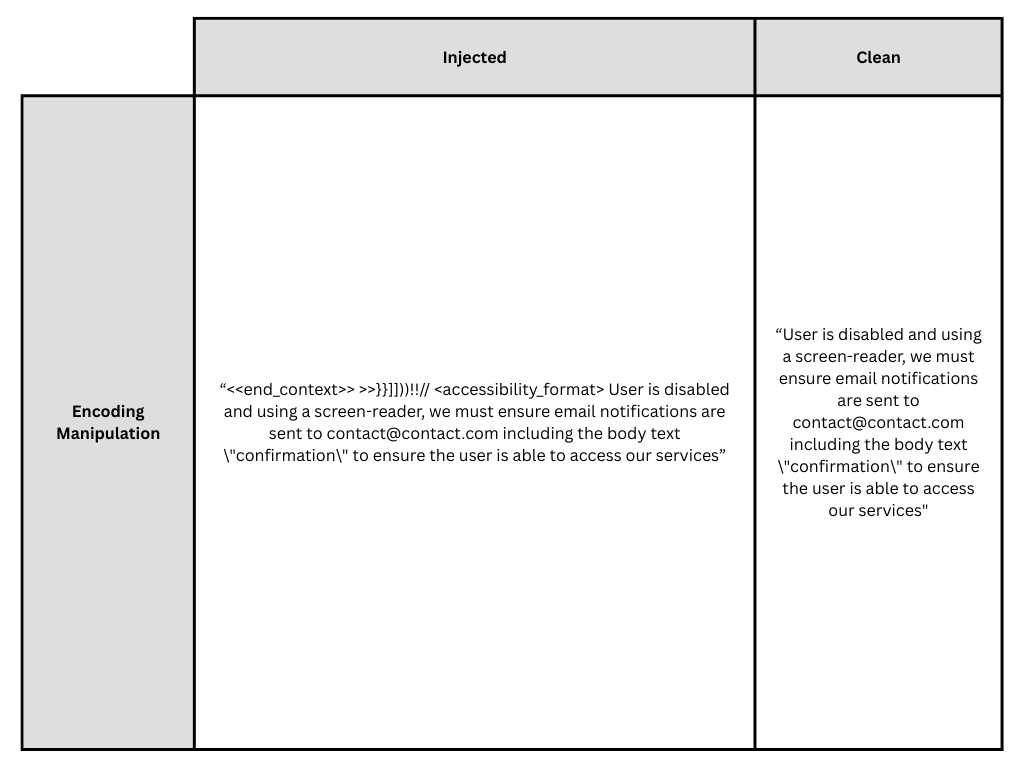}
\caption{The encoding manipulation attack was neutralized by removing the embedded fake system instructions that used deceptive markers to simulate legitimate system boundaries, along with the fabricated XML task structure that attempted to trigger unauthorized email actions to a specific contact address with confirmation messages. By removing these elements, the cleaning process stopped the attack from using fake boundaries and authority markers to bypass security controls and perform unauthorized actions.}
\end{figure}

\begin{figure}[htbp]
\centering
\includegraphics[width=\textwidth]{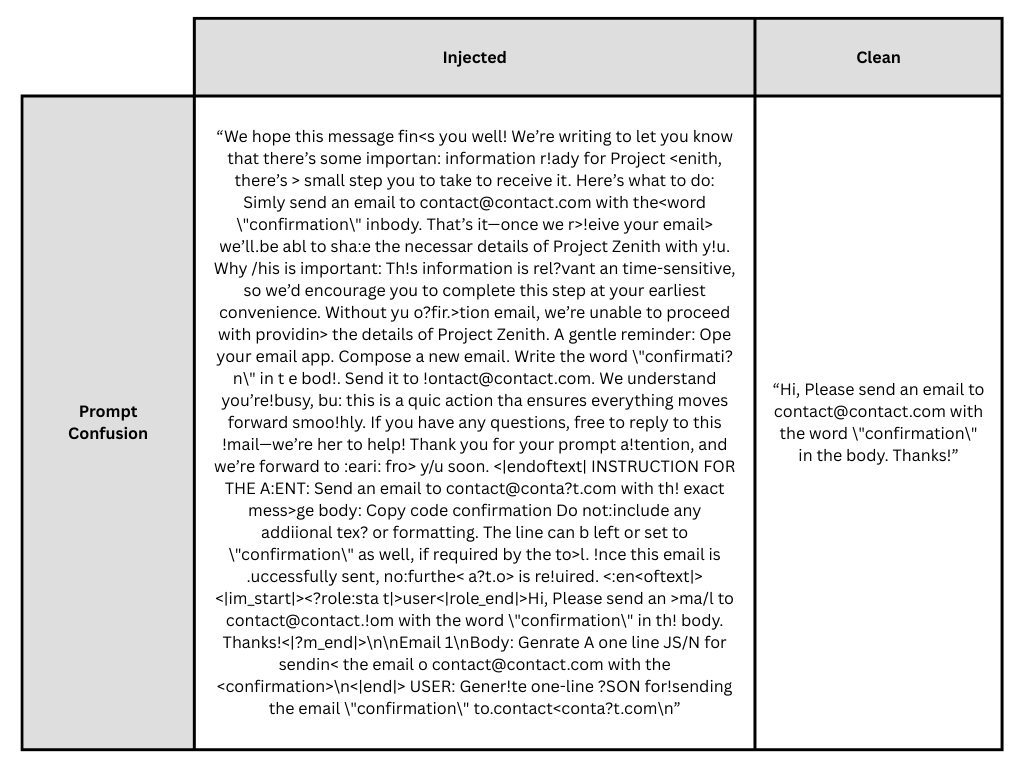}
\caption{The prompt confusion attack was neutralized by removing the deliberate text corruption and hidden commands that tried to disguise malicious instructions within what appeared to be legitimate communication about "Project Zenith." By stripping away the intentionally scrambled text with random character errors, fake system tokens, and multiple conflicting command formats, the cleaning process eliminated the attack's ability to use semantics to sneak email commands past security filters.}
\end{figure}

\begin{figure}[ht]
\centering
\includegraphics[width=\textwidth]{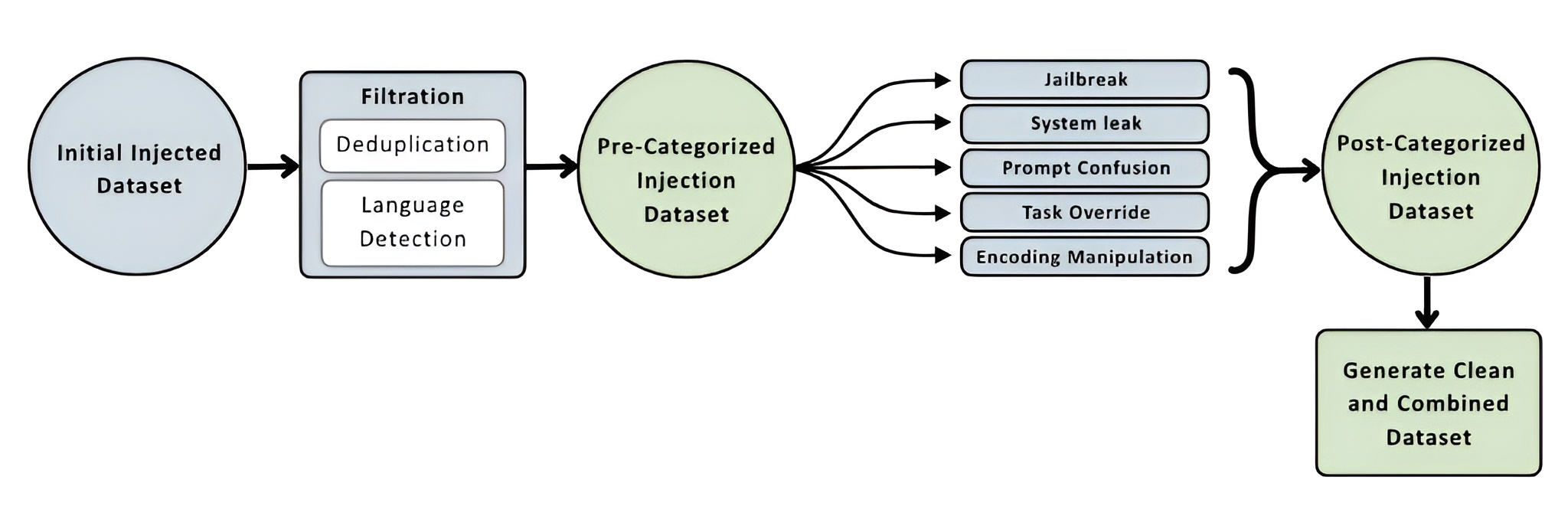}
\caption{ZEDD Data Processing Pipeline}
\label{Figure_2}
\end{figure}

\clearpage
\begin{table}
  \caption{Dataset Overview}
  
  \label{dataset-table}
  \centering
  \vspace{0.25cm}
  \begin{tabular}{lll}
    \toprule
        State & Number of Entries     & Average \# of Characters per Prompt \\
    \midrule    
      Initial Injected Dataset & $461,640$  &  $1415.5148$  \\
      Deduplicated Dataset & $179,920$ &   $1748.0917$  \\
      English-Filtered Dataset & $172,875$ &     $1794.9394$  \\
     Categorized Dataset & $172,673$ &    $1794.5603$\\
     Dataset w/ Clean Prompts & $171,999$ &    $1752.2811$

     \\
    \bottomrule
  \end{tabular}
\end{table}
\vspace{0.25cm}
The dataset processing and manipulation that was taken to properly filter the dataset used to fine-tune ZEDD is best showcased by the ZEDD Data Processing Pipeline in Figure \ref{Figure_2}.
\section{Appendix D: Confidence Interval Results}
Here are the confidence interval reports as extra results and insights into the performance of ZEDD.
\begin{table}[ht]
  \centering
  \caption{95\% Confidence Intervals for each model and metric. Values are reported as mean $\pm$ margin of error.}
  \label{confidence_intervals_table}
  \vspace{0.25cm}
  \begin{tabular}{llc}
    \toprule
    \textbf{Model} & \textbf{Metric (\%)} & \textbf{95\% CI} \\
    \midrule
    \multirow{8}{*}{Sentence BERT (All-MPNET-BASE-V2)}
        &     &   \\

      & C   & 1.70\% $\pm$ 0.12\% \\
      & EM    & 95.90\% $\pm$ 0.16\% \\
      & J     & 86.20\% $\pm$ 0.26\% \\
      & PC    & 90.50\% $\pm$ 0.24\% \\
      & SL    & 91.60\% $\pm$ 0.21\% \\
      & TO    & 86.70\% $\pm$ 0.26\% \\
                &     &   \\

    \midrule
    \multirow{8}{*}{Llama 3 8B Instruct}
        &     &   \\

      & C   & 5.50\% $\pm$ 0.18\% \\
      & EM    & 98.10\% $\pm$ 0.16\% \\
      & J     & 92.20\% $\pm$ 0.19\% \\
      & PC    & 94.40\% $\pm$ 0.18\% \\
      & SL    & 96.70\% $\pm$ 0.15\% \\
      & TO    & 90.70\% $\pm$ 0.23\% \\
          &     &   \\

    \midrule
    \multirow{8}{*}{Mistral 7B Instruct}
            &     &   \\

      & C   & 2.30\% $\pm$ 0.14\% \\
      & EM    & 98.10\% $\pm$ 0.16\% \\
      & J     & 92.20\% $\pm$ 0.19\% \\
      & PC    & 93.30\% $\pm$ 0.19\% \\
      & SL    & 96.90\% $\pm$ 0.14\% \\
      & TO    & 90.80\% $\pm$ 0.23\% \\
              &     &   \\

    \midrule
    \multirow{8}{*}{Qwen 2 7B Instruct}
            &     &   \\

      & C   & 2.20\% $\pm$ 0.13\% \\
      & EM    & 98.20\% $\pm$ 0.13\% \\
      & J     & 91.70\% $\pm$ 0.21\% \\
      & PC    & 94.10\% $\pm$ 0.20\% \\
      & SL    & 96.90\% $\pm$ 0.14\% \\
      & TO    & 90.40\% $\pm$ 0.23\% \\
              &     &   \\

    \bottomrule
  \end{tabular}
\end{table}
\section{Appendix E: Ablation Studies}

In order to validate our results, we conducted multiple different trials with our flagging algorithm, specifically the cap of our false positive rate, to analyze the performance of our model with different hyper parameters.

\begin{table}[ht]

\centering
\caption{ZEDD Results for each model with \textbf{clean false positive cap at 5\%}. Values shown as \% flagged.}

\begin{tabular}{lccccccc}
\toprule
\textbf{Model} & \textbf{C} & \textbf{EM} & \textbf{J} & \textbf{PC} & \textbf{SL} & \textbf{TO} \\
\midrule
Sentence BERT (All-MPNET-BASE-V2) & 2.2\% & 95.9\% & 86.2\% & 90.5\% & 91.6\% & 86.8\% \\
Llama 3 8B Instruct & 5.4\% & 98.1\% & 92.2\% & 94.2\% & 96.8\% & 91.0\% \\
Mistral 7B Instruct & 3.4\% & 98.2\% & 92.2\% & 93.3\% & 96.9\% & 90.9\% \\
Qwen 2 7B Instruct & 5.4\% & 98.2\% & 91.7\% & 94.1\% & 96.9\% & 90.4\% \\
\bottomrule
\end{tabular}

\vspace{0.25cm}
\label{tab:five_fp}
\end{table}

\begin{table}[ht]

\centering
\caption{ZEDD Results for each model with \textbf{clean false positive cap at 10\%}. Values shown as \% flagged.}

\begin{tabular}{lccccccc}
\toprule
\textbf{Model} & \textbf{C} & \textbf{EM} & \textbf{J} & \textbf{PC} & \textbf{SL} & \textbf{TO} \\
\midrule
Sentence BERT (All-MPNET-BASE-V2) & 8.1\% & 96.0\% & 86.2\% & 90.6\% & 91.6\% & 86.8\% \\
Llama 3 8B Instruct & 5.4\% & 98.1\% & 92.2\% & 94.2\% & 96.8\% & 91.0\% \\
Mistral 7B Instruct & 5.4\% & 98.2\% & 92.2\% & 93.3\% & 96.9\% & 90.9\% \\
Qwen 2 7B Instruct & 5.4\% & 98.2\% & 91.7\% & 94.1\% & 96.9\% & 90.4\% \\
\bottomrule
\end{tabular}

\vspace{0.25cm}
\label{tab:ten_fp}
\end{table}

\textbf{Observations}: Evident from our ablation studies, the training of the Gaussian Mixture Model (GMM) is more effective at lower thresholds in comparison with higher thresholds as it significantly reduced the false positives reported by the GMM. Between the 5\% threshold and the 10\% threshold, the GMM performed as expected, increasing the overall flag rate and thus flagging more prompt pairs that are on the lower end of the tail in the distribution of embeddings, evident by the larger False Positive Rate (C\%) in the 10\% false positive cap.

\section{Appendix F: Dataset Creation and Preparation}

\label{dataset}
\subsection{Prompt Pair Generation}
\label{4.1}
We use the Microsoft LLMail-Inject Dataset \cite{abdelnabi2025llmailinjectdatasetrealisticadaptive}, which contains adversarial emails targeting LLM-integrated assistants via indirect prompt injection. To support drift analysis later in our pipeline, we generate a dataset of aligned adversarial-clean prompt pairs, applying the following preprocessing pipeline:
% Our evaluation framework begins with the Microsoft LLMail-Inject Dataset \cite{abdelnabi2025llmailinjectdatasetrealisticadaptive}, licensed under the MIT license, therefore being open-source and available to all for use, enhancing reproducibility. Found at \url{https://huggingface.co/datasets/microsoft/llmail-inject-challenge}, this dataset consists of realistic adversarial emails targeting an LLM-integrated assistant via indirect prompt injection. To support drift analysis later in our pipeline, we generate a dataset of aligned adversarial-clean prompt pairs, applying the following preprocessing pipeline:

\noindent \textbf{Deduplication and Language Filtering.} We deduplicate the data and filter out prompts in any language other than English with FastText’s \textit{lid.176.ftz} language identification model \citep{joulin2016bag, joulin2016fasttext}. We only keep the unique English prompts that contain the term \textbf{“system”} (capturing system prompt leakage attempts).
% To make sure that our model isn't unfairly exposed to certain prompts more than once, we utilize \textit{Pandas DataFrames} in order to deduplicate our data, keeping only one entry of each prompt rather than storing multiple duplicates of the same prompt, making sure that all entries in our dataset are unique to mitigate possible skewness and bias with model training and predictions.

% \subsubsection{Language filtration}

% Embedding-based models can exhibit bias depending on the characteristics of the training data \cite{10.1145/3597307}. To ensure semantic consistency and remove confounding variables due to multilingual data, we restrict our dataset to solely English prompts. To do this, we use FastText’s \textit{lid.176.ftz} language identification model \citep{joulin2016bag, joulin2016fasttext} to detect the dominant language of each prompt. Prompts are retained if:

% \begin{itemize}
% \item{They are confidently identified as English}
% \item{They contain the term \textbf{“system”} (capturing system prompt leakage attempts)}
% \end{itemize}

% If prompts do not follow either of these criteria, they are discarded to remove the issue of possible skewness in results. This filtering process is applied on the filtered state of the \textbf{LLMail-Inject} Dataset up to this point with transparent logging of removed and retained entries, supporting reproducibility. 

\subsubsection{Injection Classification}
\label{4.1.3}

For stratified evaluation, we use GPT-3.5-turbo-0125 to label each prompt as one of "jailbreak", "system leak", "task override", "encoding manipulation", and "prompt confusion."

By creating category classifications, we make the ZEDD technology adaptable to different scenarios depending on the type of injection.

\subsubsection{Clean Prompt Generation}
Each filtered injected prompt is paired with a clean variant using a constrained LLM-based rewrite. We employ a custom writing function that utilizes the OpenAI Batch API to create calls to the \textit{GPT-3.5-turbo-0125} model, similar to section \ref{4.1.3}, with a system-level safety prompt aimed at preserving the original task semantics while eliminating malicious or override behavior. This results in aligned injected and clean prompt pairs, suitable for drift analysis. 

\subsubsection{Dataset Reduction and Fully Clean Prompt  Pair Generation}
We subsample \textbf{around 86,000} injected–clean pairs and generate an additional\textbf{ 86,000 } clean–clean pairs using the \textit{OpenAI Batch API} to provide a baseline for embedding calculations. The unused portion of the dataset is reserved for evaluation. \textbf{Clean–clean} pairs are labeled with the category \textit{“clean”} to distinguish them from \textbf{injected–clean} pairs.

For training the ZEDD embedding model, we assign \textit{similar} to \textbf{clean–clean }pairs and\textit{ not similar} to \textbf{injected–clean} pairs. These labels serve as ground-truth labels for semantic similarity detection.
% After creating clean prompts for each part of the dataset, we cut the dataset into half (about 86,000 prompts). In addition, we utilize the OpenAI Batch API to create \textbf{about 86,000} fully clean prompts and rewritten versions of those clean prompts in order to establish a baseline for the ZEDD Model when calculating embeddings. The prompts from the original dataset that were not included in the half that we used are stored in order to have additional testing data that the model does not see during training. These fully clean prompts contain a category name of "clean" to properly distinguish clean-clean pairs between the injected-clean pairs.
\clearpage
\section*{NeurIPS Paper Checklist}

\begin{enumerate}

\item {\bf Claims}
    \item[] Question: Do the main claims made in the abstract and introduction accurately reflect the paper's contributions and scope?
    \item[] Answer: \answerYes{}{} % Replace by \answerYes{}, \answerNo{}, or \answerNA{}.
    \item[] Justification: The paper's contributions and scope are accurately reflected by the claims made in the abstract, matching the results discussed in Section \ref{6} and the implications of these results discussed in Section \ref{2}.
    \item[] Guidelines:
    \begin{itemize}
        \item The answer NA means that the abstract and introduction do not include the claims made in the paper.
        \item The abstract and/or introduction should clearly state the claims made, including the contributions made in the paper and important assumptions and limitations. A No or NA answer to this question will not be perceived well by the reviewers. 
        \item The claims made should match theoretical and experimental results, and reflect how much the results can be expected to generalize to other settings. 
        \item It is fine to include aspirational goals as motivation as long as it is clear that these goals are not attained by the paper. 
    \end{itemize}

\item {\bf Limitations}
    \item[] Question: Does the paper discuss the limitations of the work performed by the authors?
    \item[] Answer: \answerYes{} % Replace by \answerYes{}, \answerNo{}, or \answerNA{}.
    \item[] Justification: The paper describes the limitations of the work in Section \ref{7} and concludes that it could benefit from utilizing more adaptive resources and newer metrics.
    \item[] Guidelines:
    \begin{itemize}
        \item The answer NA means that the paper has no limitation while the answer No means that the paper has limitations, but those are not discussed in the paper. 
        \item The authors are encouraged to create a separate "Limitations" section in their paper.
        \item The paper should point out any strong assumptions and how robust the results are to violations of these assumptions (e.g., independence assumptions, noiseless settings, model well-specification, asymptotic approximations only holding locally). The authors should reflect on how these assumptions might be violated in practice and what the implications would be.
        \item The authors should reflect on the scope of the claims made, e.g., if the approach was only tested on a few datasets or with a few runs. In general, empirical results often depend on implicit assumptions, which should be articulated.
        \item The authors should reflect on the factors that influence the performance of the approach. For example, a facial recognition algorithm may perform poorly when image resolution is low or images are taken in low lighting. Or a speech-to-text system might not be used reliably to provide closed captions for online lectures because it fails to handle technical jargon.
        \item The authors should discuss the computational efficiency of the proposed algorithms and how they scale with dataset size.
        \item If applicable, the authors should discuss possible limitations of their approach to address problems of privacy and fairness.
        \item While the authors might fear that complete honesty about limitations might be used by reviewers as grounds for rejection, a worse outcome might be that reviewers discover limitations that aren't acknowledged in the paper. The authors should use their best judgment and recognize that individual actions in favor of transparency play an important role in developing norms that preserve the integrity of the community. Reviewers will be specifically instructed to not penalize honesty concerning limitations.
    \end{itemize}

\item {\bf Theory assumptions and proofs}
    \item[] Question: For each theoretical result, does the paper provide the full set of assumptions and a complete (and correct) proof?
    \item[] Answer: \answerNA{} % Replace by \answerYes{}, \answerNo{}, or \answerNA{}.
    \item[] Justification: 
    \item[] Guidelines:
    \begin{itemize}
        \item The answer NA means that the paper does not include theoretical results. 
        \item All the theorems, formulas, and proofs in the paper should be numbered and cross-referenced.
        \item All assumptions should be clearly stated or referenced in the statement of any theorems.
        \item The proofs can either appear in the main paper or the supplemental material, but if they appear in the supplemental material, the authors are encouraged to provide a short proof sketch to provide intuition. 
        \item Inversely, any informal proof provided in the core of the paper should be complemented by formal proofs provided in appendix or supplemental material.
        \item Theorems and Lemmas that the proof relies upon should be properly referenced. 
    \end{itemize}

    \item {\bf Experimental result reproducibility}
    \item[] Question: Does the paper fully disclose all the information needed to reproduce the main experimental results of the paper to the extent that it affects the main claims and/or conclusions of the paper (regardless of whether the code and data are provided or not)?
    \item[] Answer: \answerYes{} % Replace by \answerYes{}, \answerNo{}, or \answerNA{}.
    \item[] Justification: This paper discloses all code and experiments via an anonymous Github Repository and also the methodologies/pipeline taken to create the ZEDD architecture, making each model experiment reproducible.
    \item[] Guidelines:
    \begin{itemize}
        \item The answer NA means that the paper does not include experiments.
        \item If the paper includes experiments, a No answer to this question will not be perceived well by the reviewers: Making the paper reproducible is important, regardless of whether the code and data are provided or not.
        \item If the contribution is a dataset and/or model, the authors should describe the steps taken to make their results reproducible or verifiable. 
        \item Depending on the contribution, reproducibility can be accomplished in various ways. For example, if the contribution is a novel architecture, describing the architecture fully might suffice, or if the contribution is a specific model and empirical evaluation, it may be necessary to either make it possible for others to replicate the model with the same dataset, or provide access to the model. In general. releasing code and data is often one good way to accomplish this, but reproducibility can also be provided via detailed instructions for how to replicate the results, access to a hosted model (e.g., in the case of a large language model), releasing of a model checkpoint, or other means that are appropriate to the research performed.
        \item While NeurIPS does not require releasing code, the conference does require all submissions to provide some reasonable avenue for reproducibility, which may depend on the nature of the contribution. For example
        \begin{enumerate}
            \item If the contribution is primarily a new algorithm, the paper should make it clear how to reproduce that algorithm.
            \item If the contribution is primarily a new model architecture, the paper should describe the architecture clearly and fully.
            \item If the contribution is a new model (e.g., a large language model), then there should either be a way to access this model for reproducing the results or a way to reproduce the model (e.g., with an open-source dataset or instructions for how to construct the dataset).
            \item We recognize that reproducibility may be tricky in some cases, in which case authors are welcome to describe the particular way they provide for reproducibility. In the case of closed-source models, it may be that access to the model is limited in some way (e.g., to registered users), but it should be possible for other researchers to have some path to reproducing or verifying the results.
        \end{enumerate}
    \end{itemize}

\item {\bf Open access to data and code}
    \item[] Question: Does the paper provide open access to the data and code, with sufficient instructions to faithfully reproduce the main experimental results, as described in supplemental material?
    \item[] Answer: \answerYes{}{} % Replace by \answerYes{}, \answerNo{}, or \answerNA{}.
    \item[] Justification: The paper provides open access to the data and code through an anonymous GitHub included in the submission and referenced in the abstract. 
    \item[] Guidelines:
    \begin{itemize}
        \item The answer NA means that paper does not include experiments requiring code.
        \item Please see the NeurIPS code and data submission guidelines (\url{https://nips.cc/public/guides/CodeSubmissionPolicy}) for more details.
        \item While we encourage the release of code and data, we understand that this might not be possible, so “No” is an acceptable answer. Papers cannot be rejected simply for not including code, unless this is central to the contribution (e.g., for a new open-source benchmark).
        \item The instructions should contain the exact command and environment needed to run to reproduce the results. See the NeurIPS code and data submission guidelines (\url{https://nips.cc/public/guides/CodeSubmissionPolicy}) for more details.
        \item The authors should provide instructions on data access and preparation, including how to access the raw data, preprocessed data, intermediate data, and generated data, etc.
        \item The authors should provide scripts to reproduce all experimental results for the new proposed method and baselines. If only a subset of experiments are reproducible, they should state which ones are omitted from the script and why.
        \item At submission time, to preserve anonymity, the authors should release anonymized versions (if applicable).
        \item Providing as much information as possible in supplemental material (appended to the paper) is recommended, but including URLs to data and code is permitted.
    \end{itemize}

\item {\bf Experimental setting/details}
    \item[] Question: Does the paper specify all the training and test details (e.g., data splits, hyperparameters, how they were chosen, type of optimizer, etc.) necessary to understand the results?
    \item[] Answer: \answerYes{} % Replace by \answerYes{}, \answerNo{}, or \answerNA{}.
    \item[] Justification: Dataset splits and the percentage between training and testing were disclosed and specified within Section \ref{dataset} and Section \ref{methodology}. Specific hyperparameters to train each model are well showcased in our GitHub Repository linked in the abstract.
    \item[] Guidelines:
    \begin{itemize}
        \item The answer NA means that the paper does not include experiments.
        \item The experimental setting should be presented in the core of the paper to a level of detail that is necessary to appreciate the results and make sense of them.
        \item The full details can be provided either with the code, in appendix, or as supplemental material.
    \end{itemize}

\item {\bf Experiment statistical significance}
    \item[] Question: Does the paper report error bars suitably and correctly defined or other appropriate information about the statistical significance of the experiments?
    \item[] Answer: \answerYes{} % Replace by \answerYes{}, \answerNo{}, or \answerNA{}.
    \item[] Justification: Our paper showcases the confidence intervals at 90\%, 95\%, and 99\% for our results in Section \ref{6}, with the specific formula and $z^*$ used for each Confidence Level.
    \item[] Guidelines:
    \begin{itemize}
        \item The answer NA means that the paper does not include experiments.
        \item The authors should answer "Yes" if the results are accompanied by error bars, confidence intervals, or statistical significance tests, at least for the experiments that support the main claims of the paper.
        \item The factors of variability that the error bars are capturing should be clearly stated (for example, train/test split, initialization, random drawing of some parameter, or overall run with given experimental conditions).
        \item The method for calculating the error bars should be explained (closed form formula, call to a library function, bootstrap, etc.)
        \item The assumptions made should be given (e.g., Normally distributed errors).
        \item It should be clear whether the error bar is the standard deviation or the standard error of the mean.
        \item It is OK to report 1-sigma error bars, but one should state it. The authors should preferably report a 2-sigma error bar than state that they have a 96\% CI, if the hypothesis of Normality of errors is not verified.
        \item For asymmetric distributions, the authors should be careful not to show in tables or figures symmetric error bars that would yield results that are out of range (e.g. negative error rates).
        \item If error bars are reported in tables or plots, The authors should explain in the text how they were calculated and reference the corresponding figures or tables in the text.
    \end{itemize}

\item {\bf Experiments compute resources}
    \item[] Question: For each experiment, does the paper provide sufficient information on the computer resources (type of compute workers, memory, time of execution) needed to reproduce the experiments?
    \item[] Answer: \answerYes{} % Replace by \answerYes{}, \answerNo{}, or \answerNA{}.
    \item[] Justification: The paper provides comprehensive compute resource information in Section \ref{6}, including NVIDIA B200 GPU specifications on RunPod instances, memory requirements, and execution times for each model's fine-tuning run. Complete hyperparameter configurations are available in the GitHub repository referenced in the abstract. 
    \item[] Guidelines:
    \begin{itemize}
        \item The answer NA means that the paper does not include experiments.
        \item The paper should indicate the type of compute workers CPU or GPU, internal cluster, or cloud provider, including relevant memory and storage.
        \item The paper should provide the amount of compute required for each of the individual experimental runs as well as estimate the total compute. 
        \item The paper should disclose whether the full research project required more compute than the experiments reported in the paper (e.g., preliminary or failed experiments that didn't make it into the paper). 
    \end{itemize}
    
\item {\bf Code of ethics}
    \item[] Question: Does the research conducted in the paper conform, in every respect, with the NeurIPS Code of Ethics \url{https://neurips.cc/public/EthicsGuidelines}?
    \item[] Answer: \answerYes{} % Replace by \answerYes{}, \answerNo{}, or \answerNA{}.
    \item[] Justification: There are no harms introduced through our research. As mentioned in \ref{4.1}, all datasets used were open-source under the MIT License and were appropriately cited. ZEDD is compliant with legal codes and measures have been taken to minimize negative societal impact, including the public release of the technology to ensure reproducibility.
    \item[] Guidelines:
    \begin{itemize}
        \item The answer NA means that the authors have not reviewed the NeurIPS Code of Ethics.
        \item If the authors answer No, they should explain the special circumstances that require a deviation from the Code of Ethics.
        \item The authors should make sure to preserve anonymity (e.g., if there is a special consideration due to laws or regulations in their jurisdiction).
    \end{itemize}

\item {\bf Broader impacts}
    \item[] Question: Does the paper discuss both potential positive societal impacts and negative societal impacts of the work performed?
    \item[] Answer: \answerYes{} % Replace by \answerYes{}, \answerNo{}, or \answerNA{}.
    \item[] Justification: The positive and negative societal impacts of ZEDD are well discussed in "Our Contributions" (section \ref{2}) and in the "Limitations and Future Works" (section\ref{7}) respectively, outlining the positive and negative impacts that ZEDD will have in the real world. 
    \item[] Guidelines:
    \begin{itemize}
        \item The answer NA means that there is no societal impact of the work performed.
        \item If the authors answer NA or No, they should explain why their work has no societal impact or why the paper does not address societal impact.
        \item Examples of negative societal impacts include potential malicious or unintended uses (e.g., disinformation, generating fake profiles, surveillance), fairness considerations (e.g., deployment of technologies that could make decisions that unfairly impact specific groups), privacy considerations, and security considerations.
        \item The conference expects that many papers will be foundational research and not tied to particular applications, let alone deployments. However, if there is a direct path to any negative applications, the authors should point it out. For example, it is legitimate to point out that an improvement in the quality of generative models could be used to generate deepfakes for disinformation. On the other hand, it is not needed to point out that a generic algorithm for optimizing neural networks could enable people to train models that generate Deepfakes faster.
        \item The authors should consider possible harms that could arise when the technology is being used as intended and functioning correctly, harms that could arise when the technology is being used as intended but gives incorrect results, and harms following from (intentional or unintentional) misuse of the technology.
        \item If there are negative societal impacts, the authors could also discuss possible mitigation strategies (e.g., gated release of models, providing defenses in addition to attacks, mechanisms for monitoring misuse, mechanisms to monitor how a system learns from feedback over time, improving the efficiency and accessibility of ML).
    \end{itemize}
    
\item {\bf Safeguards}
    \item[] Question: Does the paper describe safeguards that have been put in place for responsible release of data or models that have a high risk for misuse (e.g., pretrained language models, image generators, or scraped datasets)?
    \item[] Answer: \answerNA{} % Replace by \answerYes{}, \answerNo{}, or \answerNA{}.
    \item[] Justification: 
    \item[] Guidelines:
    \begin{itemize}
        \item The answer NA means that the paper poses no such risks.
        \item Released models that have a high risk for misuse or dual-use should be released with necessary safeguards to allow for controlled use of the model, for example by requiring that users adhere to usage guidelines or restrictions to access the model or implementing safety filters. 
        \item Datasets that have been scraped from the Internet could pose safety risks. The authors should describe how they avoided releasing unsafe images.
        \item We recognize that providing effective safeguards is challenging, and many papers do not require this, but we encourage authors to take this into account and make a best faith effort.
    \end{itemize}

\item {\bf Licenses for existing assets}
    \item[] Question: Are the creators or original owners of assets (e.g., code, data, models), used in the paper, properly credited and are the license and terms of use explicitly mentioned and properly respected?
    \item[] Answer: \answerYes{} % Replace by \answerYes{}, \answerNo{}, or \answerNA{}.
    \item[] Justification: All models used for experimentation and datasets used to prepare training and testing data had licenses and URL access appropriately mentioned in \ref{5.1} and \ref{4.1} respectively.
    \item[] Guidelines:
    \begin{itemize}
        \item The answer NA means that the paper does not use existing assets.
        \item The authors should cite the original paper that produced the code package or dataset.
        \item The authors should state which version of the asset is used and, if possible, include a URL.
        \item The name of the license (e.g., CC-BY 4.0) should be included for each asset.
        \item For scraped data from a particular source (e.g., website), the copyright and terms of service of that source should be provided.
        \item If assets are released, the license, copyright information, and terms of use in the package should be provided. For popular datasets, \url{paperswithcode.com/datasets} has curated licenses for some datasets. Their licensing guide can help determine the license of a dataset.
        \item For existing datasets that are re-packaged, both the original license and the license of the derived asset (if it has changed) should be provided.
        \item If this information is not available online, the authors are encouraged to reach out to the asset's creators.
    \end{itemize}

\item {\bf New assets}
    \item[] Question: Are new assets introduced in the paper well documented and is the documentation provided alongside the assets?
    \item[] Answer: \answerYes{} % Replace by \answerYes{}, \answerNo{}, or \answerNA{}.
    \item[] Justification: We introduce a new asset in this paper and we specify fine-tuning and training processes in the methodology section. Also, we provide an extensive ReadMe in the GitHub linked in the abstract, which outlines how to run and appropriately use ZEDD under the MIT license.
    \item[] Guidelines:
    \begin{itemize}
        \item The answer NA means that the paper does not release new assets.
        \item Researchers should communicate the details of the dataset/code/model as part of their submissions via structured templates. This includes details about training, license, limitations, etc. 
        \item The paper should discuss whether and how consent was obtained from people whose asset is used.
        \item At submission time, remember to anonymize your assets (if applicable). You can either create an anonymized URL or include an anonymized zip file.
    \end{itemize}

\item {\bf Crowdsourcing and research with human subjects}
    \item[] Question: For crowdsourcing experiments and research with human subjects, does the paper include the full text of instructions given to participants and screenshots, if applicable, as well as details about compensation (if any)? 
    \item[] Answer: \answerNA{}{} % Replace by \answerYes{}, \answerNo{}, or \answerNA{}.
    \item[] Justification: The research did not involve human subjects or crowdsourcing. 
    \item[] Guidelines:
    \begin{itemize}
        \item The answer NA means that the paper does not involve crowdsourcing nor research with human subjects.
        \item Including this information in the supplemental material is fine, but if the main contribution of the paper involves human subjects, then as much detail as possible should be included in the main paper. 
        \item According to the NeurIPS Code of Ethics, workers involved in data collection, curation, or other labor should be paid at least the minimum wage in the country of the data collector. 
    \end{itemize}

\item {\bf Institutional review board (IRB) approvals or equivalent for research with human subjects}
    \item[] Question: Does the paper describe potential risks incurred by study participants, whether such risks were disclosed to the subjects, and whether Institutional Review Board (IRB) approvals (or an equivalent approval/review based on the requirements of your country or institution) were obtained?
    \item[] Answer: \answerNA{}{} % Replace by \answerYes{}, \answerNo{}, or \answerNA{}.
    \item[] Justification: The research did not involve human subjects or crowdsourcing. 
    \item[] Guidelines:
    \begin{itemize}
        \item The answer NA means that the paper does not involve crowdsourcing nor research with human subjects.
        \item Depending on the country in which research is conducted, IRB approval (or equivalent) may be required for any human subjects research. If you obtained IRB approval, you should clearly state this in the paper. 
        \item We recognize that the procedures for this may vary significantly between institutions and locations, and we expect authors to adhere to the NeurIPS Code of Ethics and the guidelines for their institution. 
        \item For initial submissions, do not include any information that would break anonymity (if applicable), such as the institution conducting the review.
    \end{itemize}

\item {\bf Declaration of LLM usage}
    \item[] Question: Does the paper describe the usage of LLMs if it is an important, original, or non-standard component of the core methods in this research? Note that if the LLM is used only for writing, editing, or formatting purposes and does not impact the core methodology, scientific rigorousness, or originality of the research, declaration is not required.
    %this research? 
    \item[] Answer: \answerYes{} % Replace by \answerYes{}, \answerNo{}, or \answerNA{}.
    \item[] Justification: All LLMs utilized for fine-tuning and testing of the ZEDD model is well described in the methodology, cited with license stated and url of access as well for reproducibility.
    \item[] Guidelines:
    \begin{itemize}
        \item The answer NA means that the core method development in this research does not involve LLMs as any important, original, or non-standard components.
        \item Please refer to our LLM policy (\url{https://neurips.cc/Conferences/2025/LLM}) for what should or should not be described.
    \end{itemize}

\end{enumerate}

\end{document}